\documentclass[10pt]{article}
\usepackage{graphicx}
\usepackage{tikz}
\usetikzlibrary{arrows}
\usepackage{amsmath,amssymb}
\numberwithin{equation}{section}
\usepackage{amscd}
\usepackage{mathrsfs}
\usepackage{graphicx}
\usepackage{geometry}
\usepackage{xcolor}
\usepackage{enumitem}
\usepackage{subfig}
\usepackage{tikz}
\usetikzlibrary{arrows}
\usepackage{comment}
\usepackage{mathdots}
\usepackage{color,xcolor}
\usepackage{tikz-qtree, tikz}
\usepackage[utf8]{inputenc} 
\usetikzlibrary{decorations.pathreplacing,arrows,shapes,positioning,shadows,calc}
\usetikzlibrary{decorations, decorations.text,backgrounds}
\tikzset{every picture/.style={font issue=\footnotesize},
    font issue/.style={execute at begin picture={#1\selectfont}}
}

\newenvironment{sistema}%
{\left\lbrace\begin{array}{@{}l@{}}}%
{\end{array}\right.}
\newcommand{\pd}{probability density}

\newcommand{\diagdots}[3][-25]{%
  \rotatebox{#1}{\makebox[0pt]{\makebox[#2]{\xleaders\hbox{$\cdot$\hskip#3}\hfill\kern0pt}}}%
}
 
\begin{document}




\begin{center}
    {\large A multi-stage model of cell proliferation and death: tracking cell divisions with Erlang distributions}\\
    $~$\\
    {\small Giulia Belluccini$^1$, Mart\'in L\'opez-Garc\'ia$^a$, Grant Lythe$^a$, Carmen Molina-Par\'is$^{a,b,}$\footnote{Corresponding author: molina-paris@lanl.gov}\\
    $~$\\
{\scriptsize $^a$Department of Applied Mathematics, University of Leeds, LS2 9JT Leeds, UK\\
$^b$Theoretical Biology and Biophysics,
Theoretical Division, Los Alamos National Laboratory, Los Alamos, NM, USA \\
LA-UR-21-25591}}

\end{center}

\begin{abstract}
\noindent
Lymphocyte populations, stimulated {\em in vitro} or {\em in vivo}, grow as cells
divide. Stochastic models are appropriate because some cells undergo
multiple rounds of division, some die, and others of the same type in
the same conditions do not divide at all. If individual cells behave
independently, each can be imagined as sampling from a probability
density of times to division. The most convenient choice of density in
mathematical and computational work, the exponential density,
overestimates the probability of short division times. We consider a
multi-stage model that produces an Erlang distribution of times to
division, and an exponential distribution of times to die. The
resulting dynamics of competing fates is a type of cyton model.  Using
Approximate Bayesian Computation, we compare our model to published
cell counts, obtained after CFSE-labelled OT-I and F5 T~cells were
transferred to lymphopenic mice. The death rate is assumed to scale
linearly with the generation (number of divisions) and the number of stages
of undivided cells (generation $0$) is allowed to differ from that of
cells that have divided at least once (generation greater than zero).
Multiple stages are preferred in posterior distributions, and the mean
time to first division is longer than the mean time to subsequent
divisions. \\



\end{abstract}

\section{Introduction}
\label{intro}

Cells of the immune system patrol our bodies for months or years~\cite{westera2013closing,den2012maintenance}.
During an adaptive immune response, a subset of specific cells,
initially a small fraction of the total population, expands as cells
undergo multiple rounds of division over a few days~\cite{antia2005role}. Although most of
these cells die as the infection is overcome, lasting immunity is
ensured by the transformation, or ``differentiation'' of individual
cells to a memory phenotype.

The most convenient mathematical and computational models of the
dynamics of cell populations, 
which consider 
heterogeneity at the 
single-cell level, are Markov models.
In these models, the variables
describe the numbers of cells of each type 
as a function of time,
and cellular events such as
division, death or differentiation are
defined by their
associated rates; each event corresponds to a possible fate of an
individual cell and cells are independent of each other. In this
formulation, inter-event times are exponentially-distributed random
variables, with probability density maximised at time $0$.

Rapid expansion of cohorts of lymphocytes is recreated in
laboratories, either by stimulation {\it in vitro} or by transferring cells
to lymphopenic mice. However over the timescales of such experiments,
hours or days, it is not appropriate to treat cell division as an
instantaneous event. Rather, cells are ``cycling'' through gap,
synthesis and mitosis phases (G$_1$/G$_2$, S and M), and daughter cells cannot immediately
redivide~\cite{pandit}. To improve on the exponential distribution, Smith and Martin
proposed a model in which the time between divisions is the sum of a
fixed time spent in phase B, corresponding to S/G$_2$/M, and a variable
time spent in phase A, corresponding to G$_1$~\cite{smith-martin}. 
In the ``single stochastic division'' model of Hogan {\em et al.},
the rate of transition from A to B phase depends on the T-cell
clonotype and on the number of cells competing for the same
resources~\cite{hogan}.
If there is a common
molecular mechanism controlling the time spent in all phases of the
cell cycle, then the phase B may, instead, occupy a fixed proportion of
the total time~\cite{dowling2014stretched}.
By labelling cells with carboxyfluorescein succinimidyl ester (CFSE) or cell trace violet at the beginning of
an experiment, and then using flow cytometry at a later time, a cohort of
divided cells can be classified into generations (number of rounds of
division)~\cite{de2006estimating,gett2000cellular,hasbold1999quantitative,lee,wellard_book}. In
time-lapse microscopy experiments, individual cells are tracked and
correlations within family trees
identified~\cite{hawkins2007model,kinjyo2015real,markham2010minimum,wellard}. The cyton model is a general
framework for modelling cell cycle progression for proliferating
lymphocytes, based on the idea that each cell has a set of competing
clocks, determining its fate~\cite{hawkins2007model}.  Competition between random
times with non-exponential distributions, especially log-normal
distributions, are used to compare with data.  A number of features
are incorporated in the model: generation-dependent parameters,
heritable factors, correlations between cells of the same
generation~\cite{duffy2009impact,markham2010minimum}.

The model of Smith and Martin can be thought of as a two-stage model.
Kendall~\cite{kendall} introduced the idea of cell division
occurring at the end of a sequence of $k$ phases, with an
exponentially-distributed time spent in each phase.  Takahashi~\cite{takahashi1,takahashi2} divided
the cell cycle into four phases, with the duration of each drawn from
a Pearson type III distribution. Weber {\em et al.} postulated a
delayed exponential waiting time for each of three phases,
corresponding to G$_1$, S and G$_2$/M~\cite{weber2014quantifying}.

Here, we adopt the model of the division clock as a sequence of
phases, each with an exponential \pd, that are called stages~\cite{yates}. This
yields an Erlang distribution of times to division, while retaining
some of the mathematical and computational advantages of the
exponential distribution~\cite{yates}. The number of stages and their
mean duration can be used as free parameters to compare with
experimental data.  On the other hand, the internal stages are a
mathematical construct that do not directly correspond to biological
phases.
When the fate of an individual cell is determined by competing
internal clocks, the \pd\ of observed times between divisions is not
the same as that of the division clock because division only happens
if another fate does not. The \pd\ of division times is said to be
``censored''~\cite{duffy2012intracellular}.  When all clocks have
exponential probability densities, the \pd\ of observed division times is also
exponential due to the memoryless property of exponential random variables; this conservation of shape does not hold for
non-exponential distributions, including the log-normal and Erlang
distributions.

In this manuscript we include cell death as a competing fate in the
multi-stage approach~\cite{yates}. Assuming identical birth and death rates across
stages, we derive an analytical expression for the expected number of
cells in each stage as a function of time, and study the limiting
behaviour of the system as $t \to + \infty$. We also extend the model
by considering cell generations and tracking cell divisions across the
population of cells, in order to make theoretical predictions
comparable to CFSE experimental data~\cite{hogan}. Our multi-stage model is
a type of cyton model where the progressor fraction is set
equal to one, and where the division and death clocks are considered
to follow an Erlang and exponential distribution, respectively.
We show the applicability of our stochastic approach by calibrating
the model with CFSE data of two distinct populations of murine
T cells~\cite{hogan}. Model calibration is performed making use
of Approximate Bayesian Computation Sequential Monte Carlo (ABC-SMC)
approaches~\cite{toni}.

The paper is structured as follows. Firstly, we describe the multi-stage model and extend the approach introduced in Ref.~\cite{yates} by considering cell death in Section~\ref{model_bd}. Analytical solutions are derived under the assumption of identical birth and death rates across stages, and the limiting behaviour of the cell population is studied as $t \to + \infty$. In Section~\ref{model_gener} cell generations are introduced in the model and an analytical study of the system is carried out. In Section~\ref{results} we calibrate the multi-stage model with CFSE data from Ref.~\cite{hogan}, and compare its performance with a simple exponential model of cell division. A final discussion is provided in Section~\ref{discussion}.

\section{A multi-stage representation of cell division and death}
\label{model_bd}

\begin{figure}
    \centering
    \includegraphics[width=\textwidth]{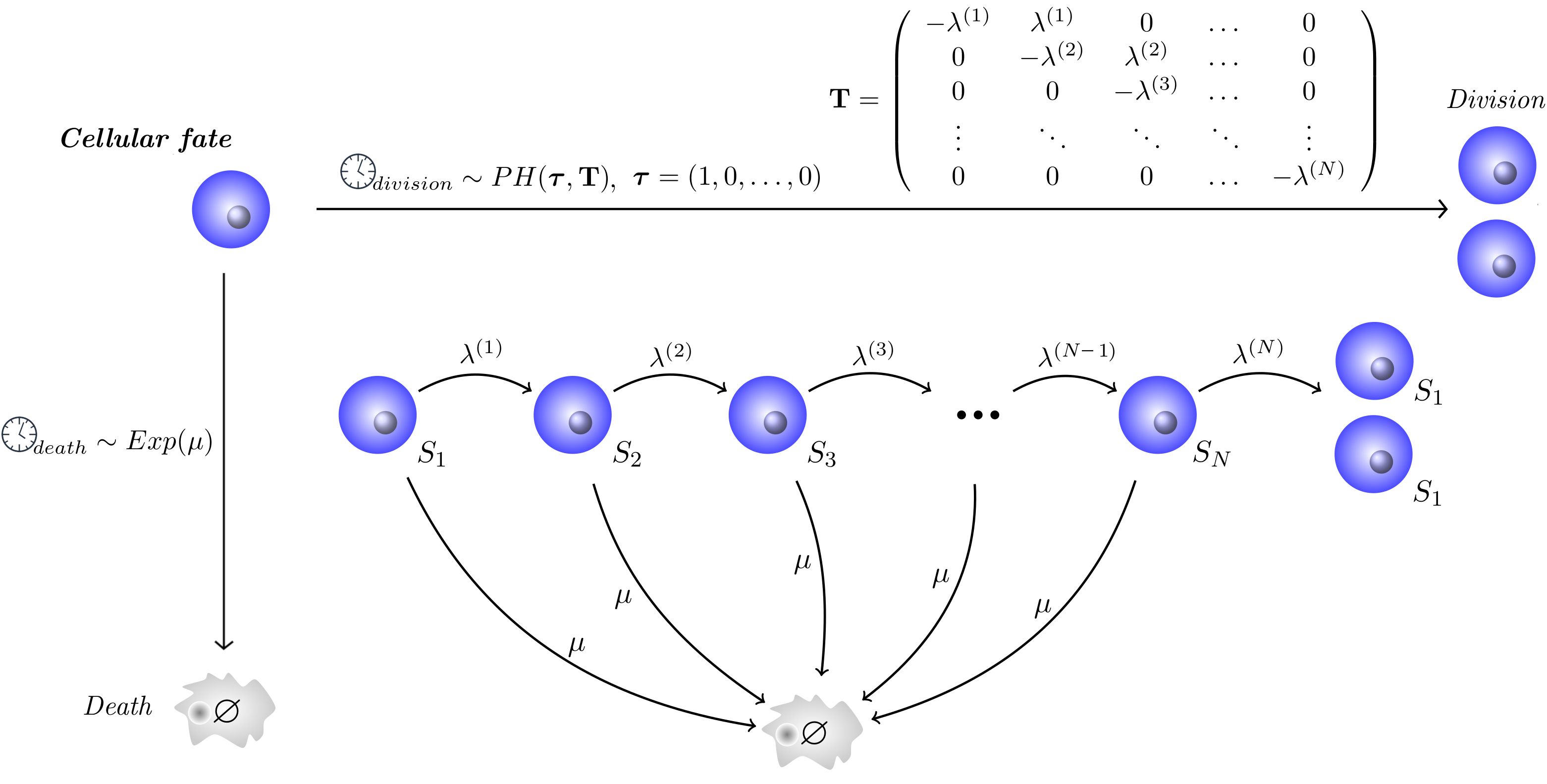}
\caption{Multi-stage model of the cell cycle. The cell cycle is divided into $N$ different stages. A  cell has to visit all the compartments (or stages) in order to divide. At each stage $j$, $j=1, \ldots,N$, the cell may proceed 
to the next stage, with birth rate $\lambda^{(j)}$, or die, with death rate $\mu$.}
\label{BeD}
\end{figure}

We present a multi-stage model of the time between cell divisions. The stages defined here are arbitrary, and not directly related to the biological phases of the cellular cycle. The cell cycle is divided into $N$ different stages, and the cell is required to 
sequentially visit 
each  compartment (or stage) in order  to divide. At each stage, each cell may either proceed to the next one or die. Let $1/\lambda^{(j)}$, $j=1,\ldots,N$, be the mean time needed to progress from stage $j$ to the subsequent one $j+1$. The time to go from stage $j$ to the next one, $j+1$, follows an exponential distribution with rate $\lambda^{(j)}$. We will refer to these rates as {\it birth} rates from now on. On the other hand, at each stage $j$ the cell may die with death rate $\mu$. Figure~\ref{BeD} shows the dynamics of the cells for the multiple stages. We note that this multi-stage representation is equivalent to considering two independent clocks for cell division and death, which compete to decide 
the cellular fate. The time-to-death clock follows an exponential distribution with rate $\mu$, while the division time follows a continuous phase-type distribution with parameters $\boldsymbol{\tau}$ and ${\bf T}$~\cite{he}, as defined in Figure~\ref{BeD}. A particular choice of phase-type distribution is the $Erlang(\lambda,N)$, which is a concatenation of $N$ identically distributed exponential steps, where all birth rates are equal: $\lambda^{(j)}=\lambda$, $j=1,\ldots,N$. We analyse this particular scenario in Section~\ref{yates_gen_identical}.

Let $S_j(t)$ be the random variable which counts the number of cells in compartment $j$ at time $t$, $j=1,\ldots,N$.
Let us denote by $M_j(t)=\mathbb{E}[S_j(t)]$, the expected value of 
$S_j(t)$. Using the moment generating function, or considering the events that can happen in a short time interval $\Delta t$ as $\Delta t \to 0^+$, the  time
 evolution of the mean number of cells in each stage can be represented 
 by the following set of differential equations
\begin{equation} \label{meantimeev_death}
\frac{dM_j(t)}{dt}=
\begin{cases}
2 \lambda^{(N)} M_N(t) -(\lambda^{(1)}+\mu) M_1(t), & \text{if} \ j=1, 
\\
\lambda^{(j-1)} M_{j-1}(t) - (\lambda^{(j)} + \mu) M_j(t), & \text{if} \ j=2, \ldots, N.
\end{cases}
\end{equation}
 We alert the reader that our analysis in the following sections is an extension of the results in Ref.~\cite{yates}, and follows similar arguments.

\subsection{Identical birth rates across stages}
\label{yates_gen_identical}

We now focus on the simpler
case where identical birth rates are assumed for each cell across different stages;
that is, $\lambda^{(j)}=\lambda,\ j=1,\ldots,N$,
so that the phase-type distribution for the time to division in Figure~\ref{BeD} is an $Erlang(\lambda,N)$.
This implies  the mean time to division is given by $\frac{N}{\lambda}$. We note here that when $N=1$, our model is nothing but a Markovian linear birth and death process with parameters $\lambda$ and $\mu$, where a cell's time to division and death are independent exponential distributions that compete to decide cellular fate. 

Under the assumption of identical birth rates across stages, Equations~\eqref{meantimeev_death} become
\begin{equation} \label{meantimeev=_death}
\frac{dM_j(t)}{dt}=
\begin{cases}
2 \lambda M_N(t) - \left(\lambda + \mu\right) M_1(t), & \text{if} \ j=1, 
\\
\lambda M_{j-1}(t) - \left(\lambda + \mu\right) M_j(t), & \text{if} \ j=2, \ldots, N.
\end{cases}
\end{equation}
 We adapt here the arguments from Ref.~\cite{yates} to analytically solve Equations~\eqref{meantimeev=_death}. In particular, we rewrite these in terms of the variables $m_j(t) = e^{(\lambda + \mu)t} M_j(t)$, $j = 1, \ldots, N$.
 This leads to
\begin{equation} \label{newvar_death}
\frac{dm_j(t)}{dt}=
\begin{cases}
2 \lambda m_N(t), & \text{if} \ j=1, 
\\
\lambda m_{j-1}(t), & \text{if} \ j=2, \ldots, N.
\end{cases}
\end{equation}
 From Equations~\eqref{newvar_death}, an $N^{th}$ order homogeneous differential equation for $m_N(t)$ follows
\begin{equation} \label{eq_to_solve_death}
\frac{d^N m_N(t)}{dt^N} = 2 \lambda^N m_N(t),
\end{equation} 
together with a set of ODEs that relate $m_j(t)$ to the derivatives of $m_N(t)$ with respect to time
\begin{equation} \label{relation_death}
m_j(t) = \left(\frac{1}{\lambda}\right)^{N-j} \frac{d^{N-j}m_N(t)}{dt^{N-j}}, \ \
j=1, \ldots, N-1.
\end{equation}
Equation~\eqref{eq_to_solve_death}
can be solved following the same arguments presented in Ref.~\cite{yates}. This leads to 
\begin{eqnarray*}
m_j(t) &=&
2^{1-\frac{j}{N}} \sum_{k=0}^{N-1} 
\; c_k \; z^{-kj} \; e^{2^{\frac{1}{N}} \lambda z^k t},
\end{eqnarray*}
 for any $j=1,\ldots,N$ and $t\geq0$, where $z=e^{\frac{2 \pi i}{N}}$ is the first $N$th root of  unity and $c_k$ (for $k=0, \ldots, N-1$) are constants which depend on the initial conditions. In particular, for $C_0$ cells in the first stage and zero cells for any other stage at time $t=0$, one gets
\begin{equation*}
c_k = \frac{C_0}{2N} \; 2^{\frac{1}{N}} \; z^k, \ \ k=0,\ldots,N-1.
\end{equation*}
Consequently, the analytical solutions of the system~\eqref{meantimeev=_death} for these initial conditions are written in terms of the original variables as
\begin{equation} \label{M_j_death}
M_j(t)=\frac{2^{\frac{1-j}{N}}C_0}{N} \; \sum_{k=0}^{N-1} 
\; z^{(1-j)k} 
e^{\left(\left(2^{\frac{1}{N}} z^k -1 \right)\lambda - \mu \right) t},
\ \ j=1,\ldots,N.
\end{equation}
Therefore, the  expected total number of cells in the population at time
$t$, $M(t)$, is computed as 
\begin{equation} \label{Mtotal_death}
M(t) = \sum_{j=1}^N M_j(t) = \frac{2^{\frac{1}{N}} C_0}{2N} 
\; \sum_{k=0}^{N-1} \; \frac{z^k}{2^{\frac{1}{N}} z^k -1} e^{\left(\left(2^{\frac{1}{N}} z^k -1\right) \lambda - \mu \right) t }.
\end{equation}

\subsection{Limiting behaviour}
\label{sec:limiting}

\begin{figure}
    \centering
    \includegraphics[width=\textwidth]{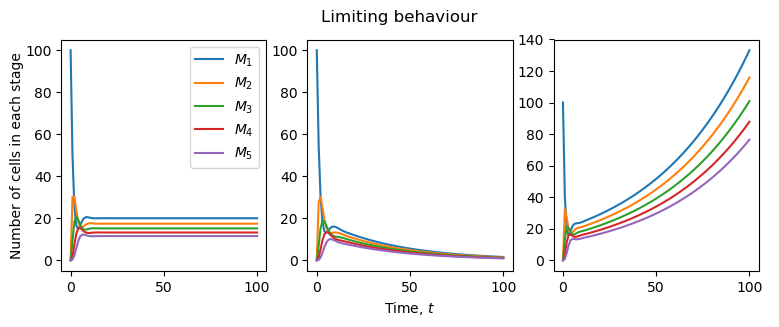}
    \caption{Limiting behaviour when $t \to +\infty$ of a population with an initial number of cells $C_0 = 10^2$. Birth and death rates, $\lambda$ and $\mu$, have units of inverse time, $t^{-1}$. {\bf Left.} Parameters: $N=5$, $\lambda = 0.6$, $\mu = (2^{1/N}-1)\lambda$. The population of cells in stage $j$ levels out to $2^{\frac{1-j}{N}} C_0/N$ for sufficiently large times.
    {\bf Centre.} Parameters: $N=5$, $\lambda=0.5$, $\mu = 0.1$. 
    The population of cells at any stage goes to extinction at late times. {\bf Right.} Parameters: $N=5$, $\lambda=0.8$, $\mu=0.1$. The populations grow according to~\eqref{Mlimit} and the relation between $M_1$ and $M_5$ given by equation~\eqref{ratio} is satisfied. For example, at $t=100$, $M_1(t) \simeq  2^{4/5}M_5(t)$.}
    \label{lim_beha}
\end{figure}

The analytical solutions above enable the study of the limiting behaviour of the population  as $t \to + \infty$, from the dual perspective of the single compartment and the  population as a whole. We first consider the coefficient of $t$ in the exponent inside the summation in Equation~\eqref{M_j_death}, to verify whether there exists a dominant term.  
When $k=0$, the exponent is given by $(2^{1/N}-1)\lambda-\mu$, which can be positive, if $\mu < (2^{1/N}-1)\lambda$ or negative, when $\mu > (2^{1/N}-1)\lambda$, or zero if $\mu = (2^{1/N}-1)\lambda$. When $k>0$, we notice that 
\begin{equation} \label{Re}
\text{Re}\left(\left(2^{\frac{1}{N}} z^k - 1 \right) \lambda - \mu \right)= \left( 2^{\frac{1}{N}} \cos{\left(\frac{2\pi k}{N}\right)} - 1 \right)\lambda - \mu.
\end{equation}
Since the cosine function is always less or equal to $1$, the right hand side of~\eqref{Re} is dominated by $\left( 2^{1/N} - 1 \right)\lambda - \mu$ for all $k = 1,\ldots,N-1$. This means that the leading term in the summation of Equation~\eqref{M_j_death} is the one corresponding to $k=0$. To conclude our analysis, we can distinguish the following three cases:

\begin{enumerate}

\item{$\mu = (2^{1/N}-1)\lambda$}. The exponent of the term corresponding to $k=0$ is zero. For $k>0$, the exponents become $2^{1/N} \lambda (z^k-1)$, which are negative for all $k=1,\ldots,N-1$. Therefore, cells at stage $j$ and the total population have the following limiting behaviour
\begin{equation*}
        \lim_{t \to + \infty} M_j(t) = \frac{2^{\frac{1-j}{N}} C_0}{N}, \ j=1,\ldots,N, \quad
        \hbox{and} 
        \quad \lim_{t \to + \infty} M(t) = \frac{2^{\frac{1}{N}} C_0}{N},
\end{equation*}
as shown in the left panel of Figure~\ref{lim_beha}.

\item{$\mu > (2^{1/N}-1)\lambda$}. The exponent of the term corresponding to $k=0$ is negative. Since it is the dominant term, the exponent is also negative for $k = 1,  \ldots,  N-1$, and therefore the cell population will become extinct as time evolves, so that $\lim_{t\to + \infty} M_j(t) = 0$ for all $j=1,\ldots,N$. Figure~\ref{lim_beha} (centre) shows an example of extinction when $N=5, \lambda=0.5 \ t^{-1}, \mu = 0.1 \ t^{-1}$ and the initial number of cells is $C_0=10^2$, where $t$ is the unit of time.

\item{$\mu < (2^{1/N}-1)\lambda$}. Since the dominant term corresponds to $k=0$,
\begin{equation} \label{lim_beha_eq}
\lim_{t \to +\infty} M_j(t) = \lim_{t \to \infty} \frac{2^{\frac{1-j}{N}} C_0}{N} e^{\left((2^{\frac{1}{N}}-1)\lambda - \mu\right)t}, \ j=1,\ldots,N.
\end{equation}
From Equation~\eqref{lim_beha_eq}, one can compute, for instance, the limiting behaviour of the ratio between $M_1(t)$ and $M_N(t)$ as
\begin{equation} 
\label{ratio}
\lim_{t \to +\infty} \frac{M_1(t)}{M_N(t)} = 2^{\frac{N-1}{N}},
\end{equation}
which is confirmed by the plot in Figure~\ref{lim_beha} (right).
To study what happens to the total population size, we compute $M(t)$ as $t \to + \infty$. We derive its time evolution  from Equation~\eqref{meantimeev=_death}
\begin{equation} \label{Mderivative_death}
\frac{dM(t)}{dt} = \lambda M_N(t) - \mu M(t).
\end{equation} 
From our previous arguments, the leading term in the summation of Equation~\eqref{Mtotal_death} is the one corresponding to $k=0$. Therefore, for large values of $t$ we have
\begin{equation} \label{Mlimit}
M(t) \simeq  
\frac{2^{\frac{1}{N}}C_0}{2N \left(2^{\frac{1}{N}}-1\right)} e^{\left(\left(2^{\frac{1}{N}} -1 \right) \lambda - \mu \right)t}.
\end{equation} 
We can notice that the expected total number of cells in the multi-stage representation is always lower than the corresponding one in the single-stage model, as it can be easily checked by considering $N=1$ in Equation~\eqref{Mlimit}.
\end{enumerate}

\subsection{Mean fraction of cells at each stage}
\label{sec:mean-proportion}

In this section we show how, analogously to the process in Yates {\em et al.}~\cite{yates} where there is no cell death, cells are not distributed proportionally to the residence time in a given stage. We define the mean fraction of cells at each stage, $P_j(t)$, as the ratio between the mean number of cells in the compartment and the expected total number of cells in the population, {\em i.e.,}
\begin{equation} \label{M_j_proportion}
P_j(t) = \frac{M_j(t)}{M(t)}, \ \ j=1,\ldots,N.
\end{equation}
From Equations~\eqref{meantimeev=_death} and \eqref{Mderivative_death}, it follows that
\begin{equation} \label{Mj_hat}
\frac{dP_j(t)}{dt}=
\begin{cases}
\lambda(2 P_N(t) - P_1(t) - P_1(t) P_N(t)), & \text{if} \ j=1, 
\\
\lambda(P_{j-1}(t) - P_j(t) - P_j(t) P_N(t)), & \text{if} \ j=2, \ldots, N,
\end{cases}
\end{equation}
which has the following steady state solution
\begin{equation} \label{ss}
P_1^*=\displaystyle{\frac{2 P_N^*}{1+P_N^*}},\quad P_j^*=\displaystyle{\frac{P_{j-1}^*}{1+P_N^*}},\  j=2, \ldots, N.
\end{equation}
One observes that $P_j^* < P_{j-1}^*$, $j=1,\ldots,N-1$, which means 
(on average) the number of cells decreases stage by stage, independently of the initial distribution of cells. In fact, one can solve Equations~\eqref{ss} to determine the analytical expression of the steady state
fraction, $P_j^*$. We have
\begin{equation}
P_j^* = \left(\sqrt[N]{2}\right)^{N-j} \left(\sqrt[N]{2} -1\right), \ \ j=1,\ldots,N,
\end{equation}
\par\noindent which does not depend on $\lambda$ or $\mu$.

\section{A multi-stage model of cell proliferation and death tracking generations}
\label{model_gener}

The mathematical model defined in Section~\ref{model_bd} is generalised here with the inclusion of cell generations. In particular, this model allows one to track the number of divisions a
given cell has undergone over time. This 
will enable one to link  cellular dynamics to CFSE data. CFSE is an intracellular dye that dilutes two-fold when a cell divides. At the beginning of the experiment  cells are labelled with the dye. Then, harvesting the cells and measuring CFSE intensity by flow cytometry at particular time instants generates cellular profiles, and by quantifying the fluorescent intensity of any given cell, one can ascertain the {\it generation} that this cell belongs to, {\em i.e.,} the number of divisions that this cell has undergone. CFSE data typically display a number of {\it intensity peaks}, which reflect the number of divisions that cells of that peak have undergone. The maximum number of peaks is usually  $9$ or $10$ due to the fact that after $10$ divisions, the intensity of the dye is $2^{10}$ fold lower than  that of the initial one, and comparable to the auto-florescence of  cells~\cite{ganusov}.

\par
\begin{figure}
\begin{center}
\includegraphics[width=\textwidth]{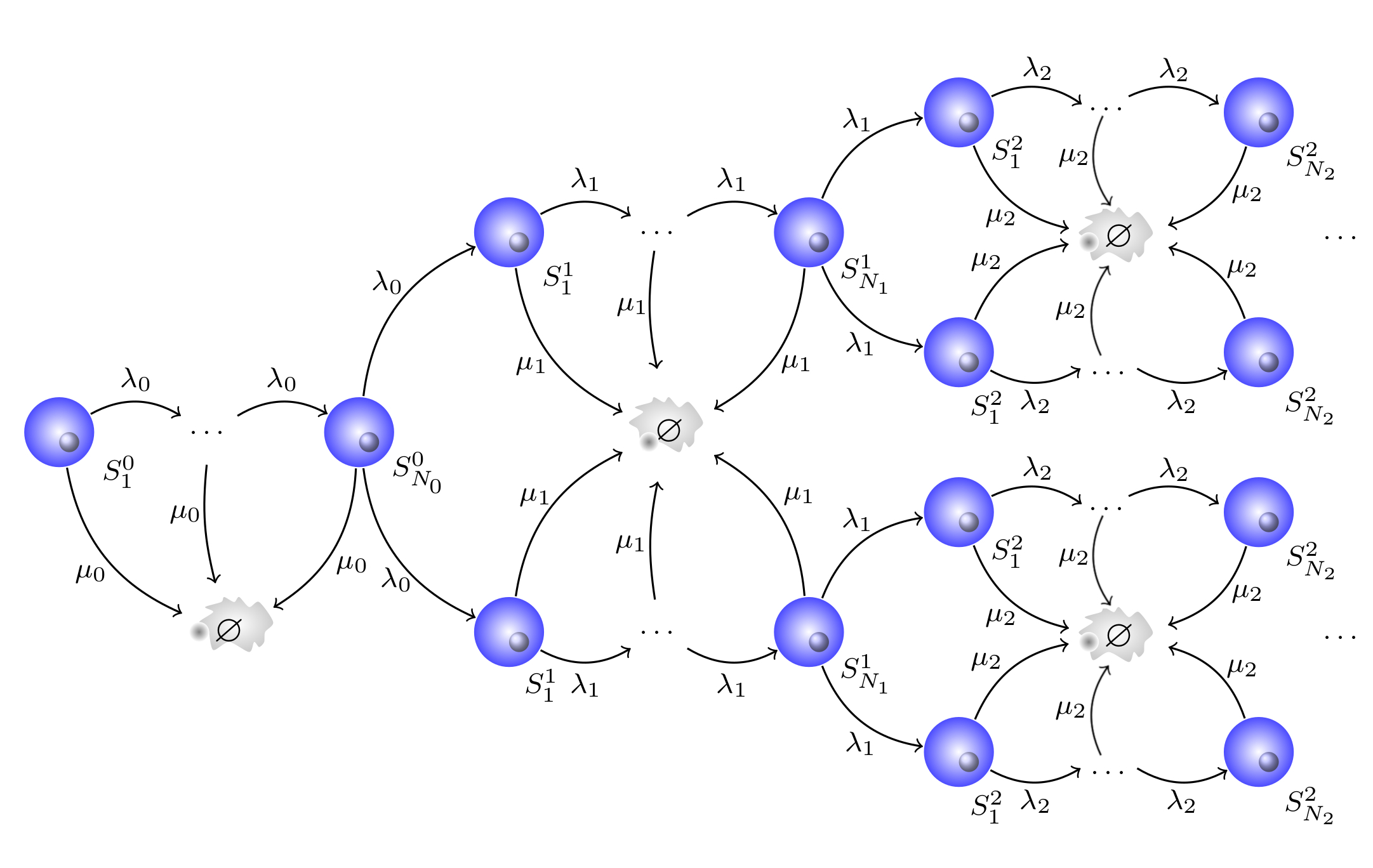}
\end{center}
\caption{Multi-stage model with cell generations. Each cell in generation $0$ has to visit all the $N_0$ compartments in order  to divide. When cells arrive 
at the last stage of generation $0$, $N_0$, they may divide with rate $\lambda_0$ or die with death rate $\mu_0$. If a cell divides, the daughter cells join the first compartment (or stage) of the next generation, and the process continues.}
    \label{model_dyn}
\end{figure}

 In our extended model, each cell cycle identifies a generation and a cell belongs to generation $g$ if it has undergone exactly $g$ divisions. Thus, cells at the beginning of the experiment, when the dye is given, would comprise generation $0$ only. Following the arguments of Section~\ref{model_bd}, and for 
a given $g$, the cell cycle is split in  $N_g$ different stages, 
where we assume this number might depend on the generation $g$ considered. A cell in generation $g$ has to sequentially visit all $N_g$ compartments to divide. On the other hand, cells might also die at each stage of the cycle. As depicted in Figure~\ref{model_dyn}, if a cell belongs to generation $g$ and lies in compartment $j$, $j=1,\ldots,N_g-1$, it may proceed to the following stage, with birth rate $\lambda_g$, or die with death rate $\mu_g$, and these rates can depend on the generation. When a cell reaches the last stage, $N_g$, of generation $g$ and divides, the two daughters will join the first compartment of generation $g+1$. In summary, given a cell in generation $g$, its time to division follows an Erlang distribution with parameters $(\lambda_g,N_g)$, whereas its time to death follows an exponential distribution with rate $\mu_g$.
These  distributions correspond to two independent competing clocks 
to control cellular fate, similarly to those considered
in Figure~\ref{BeD}.

We aim to compute the mean number of cells in each generation, which will be used to calibrate the model using CFSE data, as shown in Section~\ref{results}. We define the random variables $S^g_j(t)$, $g \ge 0$, $j=1,\ldots,N_g$, which count the number of cells in stage $j$, generation $g$, at time $t\geq0$. Letting $M^g_j(t)$ be the expected value of $S^g_j(t)$, the set of $M^g_j(t)$ obey the 
following differential equations
\begin{equation} \label{M^g_j_time}
\frac{d M^g_j(t)}{dt}=
\begin{cases}
-(\lambda_0 + \mu_0) M^0_1(t), & \text{if} \ g=0, \ j=1, \\
\lambda_g M^g_{j-1}(t) - (\lambda_g + \mu_g) M^g_j(t), & \text{if} \ g \ge 0, \ j=2, \ldots, N_g, \\
2 \lambda_{g-1} M^{g-1}_{N_{g-1}}(t) - (\lambda_g + \mu_g) M^g_1(t), & \text{if} \ g \ge 1, \ j=1.
\end{cases}
\end{equation}
In what follows, and keeping in mind 
our interest in modelling CSFE data, we assume  there exists a maximum generation $G$ that can be measured by the dye. Thus, one might be  interested in following cells from generations $g=0,\ldots,G$. For these generations, Equations~\eqref{M^g_j_time} can be solved by making use of the matrix exponential. To this end, let $\boldsymbol{M}(t)$ be the column vector of the mean number of cells in each stage and generation as time evolves, {\em i.e.,}
\begin{eqnarray*}
    \boldsymbol{M}(t) &=& \left(M^0_1(t), \dots, M^0_{N_0}(t), M^1_1(t), \dots, M^1_{N_1}(t), \dots, M^G_1(t), \dots, M^G_{N_G}(t) \right)^T\\
&=& ({\bf M}_0(t)^T,{\bf M}_1(t)^T,\dots,{\bf M}_G(t)^T)^T,    
\end{eqnarray*}
which has length $\sum_{g=0}^G N_g$, and where column sub-vectors ${\bf M}_g(t)$ contain the mean number of cells across stages in generations $g=0,\ldots,G$. Let us also define the coefficient matrix
\begin{eqnarray*}
{\bf A} &=& \left(\begin{array}{cccccc}
{\bf A}_{00} & {\bf 0}_{N_0\times N_1} & {\bf 0}_{N_0\times N_2} & \cdots & {\bf 0}_{N_0\times N_{G-1}} & {\bf 0}_{N_0\times N_{G}}\\
{\bf A}_{10} & {\bf A}_{11} & {\bf 0}_{N_1\times N_2} & \cdots & {\bf 0}_{N_1\times N_{G-1}} & {\bf 0}_{N_1\times N_{G}}\\
\vdots & \ddots & \ddots & \ddots & \vdots & \vdots\\
{\bf 0}_{N_{G-1}\times N_0} & {\bf 0}_{N_{G-1}\times N_1} & {\bf 0}_{N_{G-1}\times N_2} & \cdots & {\bf A}_{G-1,G-1} & {\bf 0}_{N_{G-1}\times N_{G}}\\
{\bf 0}_{N_{G}\times N_0} & {\bf 0}_{N_{G}\times N_1} & {\bf 0}_{N_{G}\times N_2} & \cdots & {\bf A}_{G,G-1} & {\bf A}_{G,G}\\
\end{array}\right),
\end{eqnarray*}
\par\noindent where\\
\begin{eqnarray*}
{\bf A}_{gg} &=& \left(\begin{array}{cccccc}
-(\lambda_g+\mu_g) & 0 & 0 & 0 & \cdots & 0 \\
 \lambda_g & -(\lambda_g+\mu_g) & 0 & 0 & \cdots & 0 \\
  0 & \lambda_g & -(\lambda_g+\mu_g) & 0 & \cdots & 0 \\
  \vdots  & \ddots  & \ddots & \ddots & \ddots & \vdots \\
  0  & \cdots & 0 & \lambda_g & -(\lambda_g+\mu_g) & 0\\
  0  & \cdots & 0 & 0 & \lambda_g & -(\lambda_g+\mu_g) \\
\end{array}\right),
\end{eqnarray*}
\begin{eqnarray*}
{\bf A}_{g,g-1} &=& \left(\begin{array}{cccc}
 0 & \cdots & 0 & 2 \lambda_{g-1}\\
 0 & \cdots & 0 & 0\\
 \vdots & \ddots & \vdots & \vdots \\
 0 & \cdots & 0 &  0 \\
\end{array}\right).
\end{eqnarray*}
\par\noindent ${\bf A}_{gg}$ is a square  $N_g \times N_g$ matrix, whereas ${\bf A}_{g,g-1}$ is a  $N_g \times N_{g-1}$ matrix.
$\boldsymbol{A}$ is then a real square matrix of dimension $\sum_{g=0}^G N_g$, and ${\bf 0}_{a\times b}$ represents a null matrix with dimension $a\times b$. Given the vector of the initial conditions $\boldsymbol n_0$, which has length $\sum_{g=0}^G N_g$, the system~\eqref{M^g_j_time} can be rewritten as the
following Cauchy problem
\[ 
\begin{sistema} 
    \displaystyle{\frac{d \boldsymbol{M}(t)}{dt}} = \boldsymbol{A}\cdot \boldsymbol{M}(t), \\
    \boldsymbol{M}(0) = \boldsymbol{n_0}.
\end{sistema} 
\]
The solution of the system is given by $\boldsymbol{M}(t) = e^{\boldsymbol{A} t} \boldsymbol{n_0}$, where 
\begin{equation*}
    e^{\boldsymbol{A} t} = \sum_{k=0}^{+ \infty} \frac{(\boldsymbol{A} t)^k}{k!}
\end{equation*}
\par\noindent represents the  matrix exponential . For efficient ways of computing this matrix, we refer the reader to Refs.~\cite{gomez2013maximum,gomez2014maximum,moler1978nineteen,moler2003nineteen}. Finally, we note that since CSFE data describe the number of cells in each generation, one can then compute the mean number of cells in each generation over time as 
\begin{equation}
\label{cells_in_generat}
    M^g(t) = \sum_{j=1}^{N_g} M^g_j(t), \ g \ge 0.
\end{equation}
From \eqref{M^g_j_time} and \eqref{cells_in_generat}, one can easily compute the dynamics of $M^g(t)$ as follows
\begin{equation}
\label{EQofGENS_Erlang}
    \begin{split}
        \frac{dM^0(t)}{dt} & = -\lambda_0 M^0_{N_0}(t) - \mu_0 M^0(t), \\
        \frac{dM^g(t)}{dt} & = 2 \lambda_{g-1} M^{g-1}_{N_{g-1}}(t) - \lambda_g M^g_{N_g}(t) - \mu_g M^g(t), \ g \ge 1.
    \end{split}
\end{equation}
The  solutions of the system~\eqref{M^g_j_time} can be written in a closed analytical form under more restrictive assumptions. An example is given by the hypothesis of identical number of stages and rates, $\lambda$ and $\mu$, across generations, as illustrated in detail in Section~\ref{new_gen_def}. Another simplified scenario is the one where the number of stages is equal to $1$ for all the generations, {\em i.e.,} $N_g=1$ for all $g \ge 0$, which leads to the following equations for $M^g(t)$
\begin{equation}
\begin{split}
    \label{sol_exp}
    M^0(t) & = C_0 e^{-(\lambda_0 + \mu_0)t}, \\
    M^g(t) & = 2^g\cdot C_0\cdot \left(\prod_{l=0}^{g-1} \lambda_l\right)\cdot \sum_{i=0}^g e^{-(\lambda_i + \mu_i)t} \prod_{k=0, k \neq i}^g \frac{1}{\lambda_k+\mu_k-\lambda_i-\mu_i}, \quad g\geq1.
\end{split}
\end{equation}
 Here we are considering that at time $t=0$, there are $C_0$ cells in generation $0$, so that ${\bf n}_0=(C_0,0,\ldots,0)$. In this particular case, which has been previously considered in Refs.~\cite{deboer-perelson,luzyanina2007computational,luzyanina2018,revy}, the inter-event times of cell death and division are modelled as exponential random variables, rather than Erlang distributions.

\subsection{Identical birth and death rates and number of stages across generations} 
\label{new_gen_def}

In this section we analyse the special case when the number of stages $N_g$ and the birth and death rates, $\lambda_g$ and $\mu_g$, respectively, do not depend on the generation $g$. In this case, all the cycles are comprised of the same number of stages $N$, and the common birth and death rates are denoted by $\lambda$ and $\mu$, respectively. Under these assumptions, it is possible to obtain an analytical expression for the mean number of cells in each generation. In particular, Equation~\eqref{M^g_j_time} becomes

\begin{equation} \label{M^g_j_time_id}
\frac{d M^g_j(t)}{dt}=
\begin{cases}
-(\lambda + \mu) M^0_1(t), & \text{if} \ g=0, \ j=1, \\
\lambda M^g_{j-1}(t) - (\lambda + \mu) M^g_j(t), & \text{if} \ g \ge 0, \ j=2, \ldots, N, \\
2 \lambda M^{g-1}_N(t) - (\lambda + \mu) M^g_1(t), & \text{if} \ g \ge 1, \ j=1.
\end{cases}
\end{equation}
These equations can be rewritten in terms of the new variables $m^g_j(t) = e^{(\lambda+\mu)t} M^g_j(t)$, for $g\ge0$, $j=1,\ldots,N$. Thus, Equation~\eqref{M^g_j_time_id} becomes
\begin{equation} \label{m^g_j_time_id}
\frac{d m^g_j(t)}{dt}=
\begin{cases}
0, & \text{if} \ g=0, \ j=1, \\
\lambda m^g_{j-1}(t), & \text{if} \ g \ge 0, \ j=2, \ldots, N, \\
2 \lambda m^{g-1}_N(t), & \text{if} \ g \ge 1, \ j=1.
\end{cases}
\end{equation}
To determine the solutions of Equation~\eqref{m^g_j_time_id}, we focus here on the case $M^0_1(0)=m^0_1(0)=C_0$ and all the other compartments are empty at time $t=0$. It is clear that $m^0_1(t)=C_0$ for $t\geq0$, and by solving Equation~\eqref{m^g_j_time_id} recursively one gets
\begin{equation*}
m^0_j(t) = C_0 \lambda^{j-1} \frac{t^{j-1}}{(j-1)!}, \ \ j=1, \ldots, N.
\end{equation*}
This expression allows one then to determine the mean number of cells across different stages in generation $1$, 
\begin{equation} \label{m^1_j}
m^1_j(t) = 2C_0\lambda^{N + j-1} \frac{t^{N+j-1}}{(N+j-1)!}, \ \ j=1, \ldots, N.
\end{equation}
From Equation~\eqref{m^1_j}, it follows recursively that the mean number of cells in each compartment $j$ of generation $g$ is given by
\begin{equation*}
m^g_j(t) = 2^g C_0 \lambda^{gN+j-1} \frac{t^{gN+j-1}}{(gN+j-1)!} \ \text{for} \ g \ge 0, \ j=1, \dots, N.
\end{equation*}
Going back to the original variables, $M^g_j(t)$, the solutions of \eqref{M^g_j_time_id} have the following analytical expression
\begin{equation}
    \label{M^g_j}
    M^g_j(t) = 2^g C_0 \lambda^{Ng+j-1} \frac{t^{Ng+j-1}}{(Ng+j-1)!} e^{-(\lambda+\mu)t}, \ \ g \ge 0, \ j=1,\ldots,N.
\end{equation}
As expected, we note that 
\begin{equation*}
    \lim_{t \to +\infty} M^g_j(t) = 0 \ \text{for all} \ g \ge 0, \ j=1,\ldots,N,
\end{equation*}
since cells in each generation and compartment either proceed to the next stage within their generation, divide (proceeding to the next generation), or die.

Once the mean number of cells in each compartment for a given generation is at hand, the expected number of cells in each generation can be determined according to Equation~\eqref{cells_in_generat}. In particular, we can write
\begin{equation}
\label{cell_gen_g}
    M^g(t) = \sum_{j=1}^{N} M^g_j(t) = 2^g C_0 (\lambda t)^{Ng} e^{-(\lambda+\mu)t} \sum_{j=1}^N \frac{(\lambda t)^{j-1}}{(Ng+j-1)!}, \ g \ge 0.
\end{equation}
This equation is consistent with the results of the exponential model~\cite{luzyanina2007computational}, which corresponds to the particular case $N=1$. On the other hand, if one is interested in the mean number of cells in each compartment, $M_j(t)$ for $j=1,\dots,N$, regardless of the generation that they belong to, this can be computed as
\begin{equation}
\label{stages_with_gen}
\begin{split}
M_j(t) &= \sum_{g=0}^{+\infty} M^g_j(t) = \sum_{g=0}^{+\infty} 2^g \ C_0 \ \lambda^{gN+j-1} \frac{e^{-( \lambda + \mu) t} \ t^{gN+j-1}}{(gN+j-1)!}\\
&= C_0 e^{-( \lambda + \mu) t} 2^{\frac{1-j}{N}} 
\sum_{g=0}^{+\infty} \frac{\left(2^{\frac{1}{N}} \lambda t\right)^{gN+j-1}}{(gN+j-1)!},
\end{split}
\end{equation}
\par\noindent for $j=1,\ldots,N$ and $t\geq0$. In practice, one could truncate the series above to get an approximation of the mean number of cells in each stage. However, we note that one can use instead the solution provided by Equation~\eqref{M_j_death}, since the dynamics of this model is  equivalent to that described in Section~\ref{model_bd}, when the parameters $N$, $\lambda$ and $\mu$ are generation-independent. It can be  numerically checked, that this indeed provides equivalent results. In fact, when $N=1$ or $N=2$, it is straightforward to analytically prove the equivalence. In the former case, it is enough to recall the power series of the exponential function. In the latter case, we derive from \eqref{M_j_death}
\begin{eqnarray*}
    M_1(t) &=& \frac{C_0}{2} e^{-(\lambda+\mu)t} \left( e^{\sqrt{2} \lambda t} + e^{- \sqrt{2} \lambda t} \right),
    \\
    M_2(t) &=&  \frac{C_0}{2 \sqrt{2}} e^{-(\lambda+\mu)t} \left( e^{\sqrt{2} \lambda t} - e^{- \sqrt{2} \lambda t} \right),
\end{eqnarray*}
where we used the fact that $z=e^{\pi i}=-1$. On the other hand, from \eqref{stages_with_gen} we obtain
\begin{eqnarray*}
    M_1(t) &=& C_0 e^{-(\lambda+\mu)t} \cosh \left(\sqrt{2} \lambda t\right)
    = C_0 e^{-(\lambda+\mu)t} \frac{e^{\sqrt{2} \lambda t} + e^{- \sqrt{2} \lambda t}}{2}, \\
    M_2(t) &=& C_0 e^{-(\lambda+\mu)t} \sinh \left(\sqrt{2} \lambda t\right)
    = \frac{C_0}{\sqrt{2}} e^{-(\lambda+\mu)t} \frac{e^{\sqrt{2} \lambda t} - e^{- \sqrt{2} \lambda t}}{2},
\end{eqnarray*}
which shows that the two models lead to the same result for the expected number of cells in each stage.

\subsection{Comparison with the cyton model}
\label{Cyton_model}

The {\em cyton model} is a stochastic model proposed to describe the population dynamics of B and T lymphocytes~\cite{hawkins2007model}. Division and death times are regulated by two independent clocks, and the competition between these mechanisms determines the fate of the cell (see Figure~\ref{BeD}). When a cell divides, these clocks, which depend on the number of divisions the cell has undergone, are reset for each daughter cells. However, when analysing an {\it in vitro} experiment with this type of cells, there is evidence that not all cells either divide or die. For instance, a portion of them may not respond to the stimulation~\cite{pereira}, or may respond without division~\cite{deenick}. This is the reason why a progressor fraction is defined in the cyton model.  This progressor fraction represents
for a given generation, the fraction of cells that are capable of undergoing further division. Each cellular fate mechanism is described in terms of a probability density function, and the parameters that define these probabilities are the free parameters in the model. Right skewed distributions, such as log-normal or gamma, are usually adopted to characterise the 
two independent clocks that regulate cell division and death. 
In summary, the cyton model is based on the following assumptions:
\begin{itemize}
    \item death and division are stochastic processes, characterised by a probability density function for the time to divide or die, respectively,
    \item these processes are independent, and compete to determine the fate of the cell,
    \item the clocks responsible for these processes are reset when a cell divides,
    \item only a fraction of the cells in each generation are capable to undergo further divisions, and 
    \item the machineries that regulate cellular fate depend on the cell's generation.
\end{itemize}
In order to translate these assumptions into mathematical terms, let $\gamma_g$ be the progressor fraction characterising cells having undergone $g$ divisions, and let $\phi_g(\cdot)$ and $\psi_g(\cdot)$ represent the probability density functions for the time to division and death, respectively, for cells in generation $g$. As described in Ref.~\cite{hawkins2007model}, the number of cells dividing for the first time, or dying, per unit time at time $t\geq 0$ can be calculated, respectively, as:
\begin{align}
    \label{n^div_0}
    n^{div}_{0}(t) & = \gamma_0 \; C_0 \cdot \left(1-
    \int_0^t \psi_0(s) \; ds \right) \cdot \phi_0(t), \\
    \label{n^die_0}
    n^{die}_{0}(t) & = C_0 \cdot \left(1-
    \gamma_0 \int_0^t \phi_0(s) \; ds \right) \cdot \psi_0(t),
\end{align}
where $C_0$ is the initial number of cells in the population. Consequently, the time evolution of the expected number of cells in generation $0$, ${\widetilde M}^0(t)$, obeys the differential equation 
\begin{equation}
    \label{Cyton_0}
    \frac{d{\widetilde M}^0(t)}{dt} = -\left[n^{div}_{0}(t) + n^{die}_{0}(t)\right].
\end{equation}
On the other hand, the number of cells in generation $g$ dividing, or dying, per unit time at time $t$ can be computed, respectively, as
\begin{align}
    \label{n^div_g}
    n^{div}_{g}(t) & = 2 \gamma_g \int_0^t n^{div}_{g-1}(s) \cdot \left[1-
    \int_0^{t-s} \psi_g(l) \; dl \right] \cdot \phi_g(t-s) \; ds, \\
    \label{n^die_g}
    n^{die}_{g}(t) & = 2 \int_0^t n^{div}_{g-1}(s) \cdot \left[1-
    \gamma_g \int_0^{t-s} \phi_g(l) \; dl \right] \cdot \psi_g(t-s) \; ds.
\end{align}
Hence, the dynamics of the average number of cells in each generation as time evolves, ${\widetilde M}^g(t)$, is governed by the differential equations
\begin{equation}
\label{Cyton_eqns}
    \frac{d{\widetilde M}^g(t)}{dt} = 2 n^{div}_{g-1}(t) - n^{div}_{g}(t) - n^{die}_{g}(t), \ g \ge 1.
\end{equation}
 In the next sections we show how the cyton model is equivalent to our model for particular choices of the probability density functions of the division and death clocks, $\phi_g(\cdot)$ and $\psi_g(\cdot)$, and the progressor faction $\gamma_g$.

\subsubsection{Exponential time to division and death} 

 We now assume that  the number of stages in all the generations is equal to one, {\em i.e.,} $N_g =1$ for all $g \ge 0$.
 This means that cells in generation $g$ divide after an exponentially distributed time with rate $\lambda_g$, and die with rate $\mu_g$. Therefore, Equations~\eqref{EQofGENS_Erlang} become 
\begin{equation}
\label{EQofGENS_exp}
    \begin{split}
        \frac{dM^0(t)}{dt} & = -\left(\lambda_0+\mu_0\right) M^0(t), \\
        \frac{dM^g(t)}{dt} & = 2 \lambda_{g-1} M^{g-1}(t) - \left(\lambda_g+\mu_g\right) M^g(t), \ g \ge 1.
    \end{split}
\end{equation}
It is clear that, in this case, our model is equivalent to the cyton model with exponential times for division and death, and progressor fraction $\gamma_g=1$, $g\geq 0$. One can show this analytically by proving that $n^{div}_{g}(t) = \lambda_g M^g(t)$ and $n^{die}_{g}(t) = \mu_g M^g(t)$.
This can be shown by induction on $g$. In the cyton model, the assumption of exponential time to division and death implies that $\phi_g(t) = \lambda_g e^{-\lambda_g t}$ and $\psi_g(t) = \mu_g e^{-\mu_g t}$, $g \ge 0$. Therefore, according to Equations~\eqref{n^div_0} and \eqref{n^die_0}, the number of cells dividing for the first time or dying to exit generation $0$ per unit time at time $t$ is
\begin{equation*}
    n^{div}_{0}(t) \ =\ C_0  \lambda_0 e^{-(\lambda_0+\mu_0) t}, \quad     n^{die}_{0}(t) \ = \ C_0 \mu_0 e^{-(\lambda_0+\mu_0) t}.
\end{equation*}
From \eqref{sol_exp} we know that in our model $M^0(t) = C_0 e^{-(\lambda_0+\mu_0) t}$. Therefore, $n^{div}_{0}(t)= \lambda_0 M^0(t)$ and $n^{die}_{0}(t)= \mu_0 M^0(t)$, which proves the case $g=0$. 
We assume the identities $n^{div}_{g}(t) = \lambda_g M^g(t)$ and $n^{die}_{g}(t) = \mu_g M^g(t)$ hold for generation $g$ and we prove them for generation $g+1$. From \eqref{sol_exp} and \eqref{n^div_g}, we can write
\begin{align*}
    n^{div}_{g+1}(t) & = 
    2 \int_0^t \lambda_g 
    2^g C_0 \prod_{l=0}^{g-1} \lambda_l \sum_{i=0}^g e^{-(\lambda_i + \mu_i)s} \prod_{k=0, k \neq i}^g \frac{1}{\lambda_k+\mu_k-\lambda_i-\mu_i}
    \lambda_{g+1} e^{-(\lambda_{g+1} + \mu_{g+1})(t-s)} ds \\
    & = \lambda_{g+1} 2^{g+1} C_0 \prod_{l=0}^g \lambda_l \sum_{i=0}^g 
    e^{-(\lambda_{g+1} + \mu_{g+1})t} \int_0^t 
    \prod_{k=0, k \neq i}^g
    \frac{e^{(\lambda_{g+1} + \mu_{g+1} - \lambda_g -\mu_g)s}}{\lambda_k+\mu_k-\lambda_i-\mu_i} ds \\
    & = \lambda_{g+1} M^{g+1}(t).
\end{align*}
For the number of cells in generation $g+1$ dying, Equation~\eqref{n^die_g}, together with Equation~\eqref{sol_exp} lead to
\begin{align*}
    n^{die}_{g+1}(t) & =
    2 \int_0^t \lambda_g 
    2^g C_0 \prod_{l=0}^{g-1} \lambda_l \sum_{i=0}^g e^{-(\lambda_i + \mu_i)s} \prod_{k=0, k \neq i}^g \frac{1}{\lambda_k+\mu_k-\lambda_i-\mu_i}
    \mu_{g+1} e^{-(\lambda_{g+1} + \mu_{g+1})(t-s)} ds \\
    & = \mu_{g+1} 2^{g+1} C_0 \prod_{l=0}^g \lambda_l \sum_{i=0}^g 
    e^{-(\lambda_{g+1} + \mu_{g+1})t} \int_0^t 
    \prod_{k=0, k \neq i}^g
    \frac{e^{(\lambda_{g+1} + \mu_{g+1} - \lambda_g -\mu_g)s}}{\lambda_k+\mu_k-\lambda_i-\mu_i} ds \\
    & = \mu_{g+1} M^{g+1}(t),
\end{align*}
which concludes the proof. Making use of the identities $n^{div}_{g}(t) = \lambda_g M^g(t)$ and $n^{die}_{g}(t) = \mu_g M^g(t)$ in \eqref{Cyton_0} and \eqref{Cyton_eqns}, one obtains that $M^g(t)$ and ${\widetilde M}^g(t)$ obey the same differential equations for all $g \ge 0$. Thus, the two models are equivalent.

\subsubsection{Erlang time to division and exponential time to death} 

We now consider the more interesting scenario where the number of stages in each generation is greater than one, so that one gets a multi-stage representation for the proliferation process of each cell. We focus here on the scenario described in Section~\ref{new_gen_def}, where identical number of stages $N$ and birth and death rates, $\lambda$ and $\mu$, respectively, are considered across generations. Similarly to the previous case, we prove that $n^{div}_{g}(t) = \lambda M^g_N(t)$ and $n^{die}_{g}(t) = \mu M^g(t)$ by induction on $g$. Since a cell's time to division is Erlang distributed and a cell's time to death is exponentially 
distributed, $\psi_g(t) = \mu e^{-\mu t}$ for all $g \ge 0$ and 
\begin{equation*}
    \phi_g(t) = \frac{\lambda^N t^{N-1} e^{-\lambda t}}{(N-1)!}, \ g \ge 0,
\end{equation*}
where the progressor fraction is again set to $1$ for each generation. Note that in this case the parameters in $\phi_g(\cdot)$ and $\psi_g(\cdot)$ are independent of the generation $g$, since the number of stages and the birth and death rates are identical across generations. From \eqref{n^div_0} and \eqref{n^die_0}, the number of cells dividing for the first time or dying to exit generation $0$  per unit time at time $t$ is
\begin{equation*}
    n^{div}_{0}(t) \ =\ \frac{C_0 \lambda^N t^{N-1}}{(N-1)!} e^{-(\lambda+\mu) t}, \quad
    n^{die}_{0}(t) \ =\ C_0 \mu e^{-(\lambda+\mu) t} \sum_{j=0}^{N-1} \frac{(\lambda t)^j}{j!}.
\end{equation*}
The dynamics of the expected number of cells in generation $0$  is given by \eqref{Cyton_0}, as in the previous case. From \eqref{M^g_j} and \eqref{cell_gen_g}, we know that in our model, we have
\begin{equation*}
        M^0(t) \ =\ C_0 e^{-(\lambda+\mu) t} \sum_{j=0}^{N-1} \frac{(\lambda t)^j}{j!}, \quad
        M^0_N(t) \ =\ \frac{\lambda^N t^{N-1}}{(N-1)!} e^{-(\lambda+\mu) t}.
\end{equation*}
Therefore, $n^{div}_{0}(t)= \lambda M^0_N(t)$ and $n^{die}_{0}(t)= \mu M^0(t)$, which concludes the case $g=0$. Replacing these identities in \eqref{Cyton_0} leads to 
\begin{equation*}
    \frac{d{\widetilde M}^0(t)}{dt} = -\lambda M^0_N(t) - \mu M^0(t),
\end{equation*}
which is the differential equation derived in \eqref{EQofGENS_Erlang} for $M^0(t)$ in our model. Now, we suppose that the identities $n^{div}_{g}(t) = \lambda M^g_N(t)$ and $n^{die}_{g}(t) = \mu M^g(t)$ hold for generation $g$ and we prove them for generation $g+1$. From \eqref{n^div_g} and the induction hypothesis,
\begin{align*}
    n^{div}_{g+1}(t) & = 2 \int_0^t \lambda 2^g C_0 
    \frac{(\lambda s)^{Ng + N -1}}{(Ng + N -1)!} e^{-(\lambda + \mu)s} e^{-\mu(t-s)} 
    \frac{\lambda^N (t-s)^{N-1} e^{-\lambda(t-s)}}{(N -1)!} ds \\
    & = 2^{g+1} \frac{\lambda^{N(g+2)}}{(N(g+1)-1)!} C_0 e^{-(\lambda + \mu)t} \frac{1}{(N-1)!} \int_0^t s^{N(g+1)-1} (t-s)^{N-1} ds \\
    & = 2^{g+1} \frac{\lambda^{N(g+2)}}{(N(g+1)-1)!} C_0 e^{-(\lambda + \mu)t}
    \sum_{j=0}^{N-1} \frac{(-1)^j t^{N-1-j}}{j! (N-1-j)!} 
    \int_0^t s^{N(g+1)-1+j}ds \\
    & = \lambda 2^{g+1} \frac{(\lambda t)^{N(g+1)+N-1}}{(N(g+1)+N-1)!} C_0 e^{-(\lambda + \mu)t} \ =\ \lambda M^{g+1}_N(t),
\end{align*}
where we used Equation~\eqref{M^g_j} for the last identity.

If we now look at the
 number of cells in generation $g+1$ dying per unit of time, Equation~\eqref{n^die_g}, together with the induction hypothesis, leads to
\begin{align*}
    n^{die}_{g+1}(t) & = 2 \int_0^t \lambda 2^g C_0 
    \frac{(\lambda s)^{Ng + N -1}}{(Ng + N -1)!} e^{-(\lambda + \mu)s} e^{-\lambda(t-s)} 
    \sum_{j=0}^{N-1} \frac{\lambda^j (t-s)^j}{j!}
    \mu e^{-\mu(t-s)} ds \\
    & = 2^{g+1} \lambda^{Ng + N} C_0 \frac{e^{-(\lambda + \mu)t} \mu}{(Ng + N -1)!} 
    \sum_{j=0}^{N-1} \frac{\lambda^j}{j!} \int_0^t s^{Ng + N - 1} (t-s)^j ds \\
    & = 2^{g+1} \lambda^{Ng + N} C_0 e^{-(\lambda + \mu)t} \mu 
    \sum_{j=0}^{N-1} \lambda^j 
    \sum_{k=0}^j \frac{t^j}{k!(j-k)!} \frac{t^{k+N+Ng}}{k+N+Ng} 
    \frac{(-1)^k}{(Ng+N-1)!} \\
    & = \mu 2^{g+1} C_0 e^{-(\lambda + \mu)t} \sum_{j=0}^{N-1} \frac{(\lambda t)^{N(g+1)+j}}{(N(g+1)+j)!} \ = \ \mu M^{g+1}(t),
\end{align*}
where the last identity was obtained making
use of Equation~\eqref{cell_gen_g}.
Hence, Equation~\eqref{Cyton_eqns} becomes
\begin{equation*}
    \frac{d{\widetilde M}^g(t)}{dt} = 2 \lambda M^{g-1}_N(t) - \lambda M^g_N(t) - \mu M^g(t), \ g \ge 1,
\end{equation*}
which is identical to \eqref{EQofGENS_Erlang} for $M^g(t)$, $g \ge 1$.
This concludes the proof of  the equivalence between the cyton model and our model with generations when a cell's time to divide is Erlang distributed with parameters $\lambda$ and $N$, and a cell's time to death is assumed to be an exponential with rate $\mu$. 
In this way we have shown that the analysis presented in this section
for  the multi-stage model with Erlang division time and exponential death time
leads to exact closed solutions for the cyton model
with the previous choice of clocks.

\section{Model calibration}
\label{results}

In this section we illustrate how the multi-stage model tracking cell generations can be calibrated making use of CFSE data. To this end 
we perform Approximate Bayesian Computation based on Sequential Monte Carlo (ABC-SMC) methods~\cite{toni}.
Parameter inference is performed with the multi-stage model with cell generations described in Section~\ref{model_gener}, and its exponential (or single-stage) version, which results from setting the number of stages equal to $1$
for all generations, {\em i.e.,} $N_g = 1$, $g\geq0$. 

\begin{figure}[htbp]
    \centering
    \includegraphics[width=\textwidth]{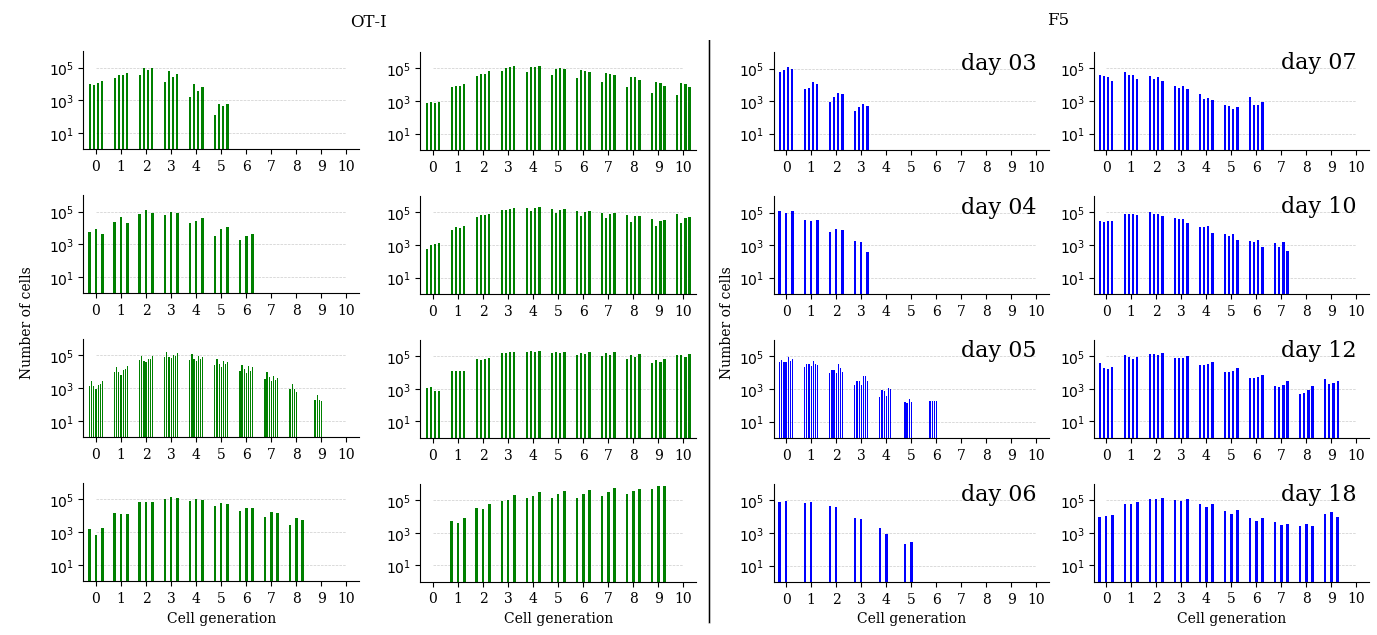}
    \caption{Data set of murine T~lymphocytes.
    Left: OT-I T~cells.   Right: F5 lymphocytes. For each time point, the number of cells is plotted for each mouse and  generation.}
    \label{hogan_dataset}
\end{figure}

The data sets we make use of are taken from an experimental study of lymphopenia-induced proliferation~\cite{hogan}. This response has been observed to vary between different T~cell clonotypes ({\em i.e.,} the set of T~cells with the same T~cell receptor (TCR) expressed on their surface). Hogan {\rm et al.} 
transferred CFSE-labelled OT-I or F5 T~cells
 intravenously to lymphopenic mice. A certain number of days (3, 4, 5, 6, 7, 10, 12 and 18 days) after the transfer, spleens and lymph nodes were recovered from the mice and analysed by flow cytometry to quantify the expression levels of CD8, CD5, CD44, and CFSE dilution~\cite{hogan}. 
 For each time point the number of mice analysed was 
between 3 and 7. We note that two independent transfer experiments,
carried out under identical conditions, 
were performed: one for OT-I cells and a second one for F5 (see 
 Figure~\ref{hogan_dataset}). 
 
\begin{figure}[h!]
    \centering
    \includegraphics[scale=0.6]{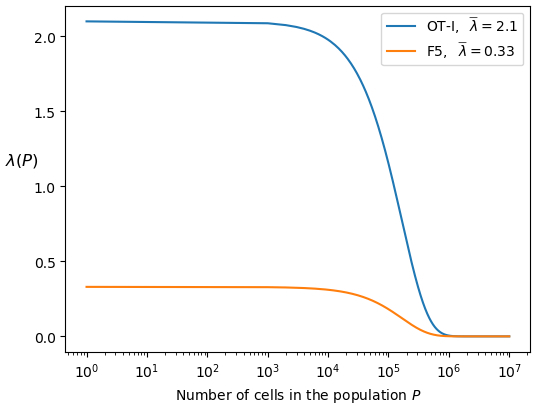}
    \caption{Density-dependent birth rate, $\lambda(P)$, as a function of the population size, $P$. The parameter ${\bar \lambda}$, with units of  $cell \cdot day^{-1}$, represents the rate of growth under no competition and $\delta$ quantifies the level of reduction caused by the expansion of competing cells. Values for ${\bar \lambda}$ (shown in the inset)
    and $\delta=6.0 \times 10^{-6}$ are taken from  \cite[Table 1]{hogan}.}
    \label{lambda_P}
\end{figure}

 Figure~\ref{hogan_dataset} clearly
 shows that OT-I T~cells 
 proliferate faster than F5,
 so that by day 7 there are OT-I cells in generation 10,
 whereas for F5 cells the maximum generation at day 7 is 6.
 This greater proliferation of OT-I cells eventually leads, after one week, to competition for resources ({\em e.g.,} IL-7 cytokine) and 
 the OT-I population approaching a carrying capacity~\cite{hogan}. 
 Since our model does not account for competition, 
 it can only appropriately describe the  dynamics 
 of OT-I cells during the first week of the experiment.
 Thus, 
  for OT-I cells 
 we will only make use of the data set until day 7.
 Yet for the F5 population we will use the entire data set.
  In Ref.~\cite{hogan} this competition was explained  with the assumption of a density-dependent birth rate, $ \lambda(P)$, as follows 
\begin{equation}
\label{lam_eq}
    \lambda(P) = {\bar \lambda} \; e^{-\delta P},
\end{equation}
where ${\bar \lambda}$ represents the rate of growth in the condition of unlimited resources, $\delta$ quantifies the size of reduction caused by the expansion of competing cells, and $P$ is the size of the population~\cite{hogan}. Figure~\ref{lambda_P} 
shows the density-dependent birth rate, $\lambda(P)$, as a function of the population size $P$. It  suggests that the competition for resources is more significant in the case of OT-I T~cells.
In the experiments
 the number of OT-I cells after one week (about $5 \times 10^5$) is larger than the population of F5 T~cells at day 18 (about $4 \times 10^5$). Therefore, the population of F5 T~lymphocytes never reaches the carrying capacity and the role of competition for resources can be neglected for this clonotype.
Before performing the Bayesian inference, we make some assumptions in our model. Several studies  have shown that the first division in this type of experiments usually requires a longer time than subsequent\footnote{We note that the term {\em subsequent division} denotes any cell division event but the first.} divisions need, since cells may take some time to become activated before starting to divide~\cite{hawkins2007model,kinjyo2015real,markham2010minimum}. Thus, we assume here that all generations but $0$ are comprised of the same number of stages $N$, whereas generation $0$ is characterised by $N_0$ stages. Similarly, cells in generation $0$ proceed to divide with birth rate $\lambda_0$, whilst all the other generations have a birth rate $\lambda$. On the other hand, we propose 
per cell death rates over generations to be linearly dependent on the number of cell divisions that the cell has undergone~\cite{ganusov,mazzocco},  as follows
\begin{equation}
    \label{death_rates}
    \mu_g = \alpha \cdot g, \ g \ge 0,
\end{equation}
where $\alpha$ is a parameter to estimate. These linear death rates 
encode the fact that cells are more likely to die when they have already undergone several divisions~\cite{ganusov,mazzocco}. We also merge, for the dataset in 
Figure~\ref{hogan_dataset}, cells within the highest generations into a single group $5+$, which combines all the cells that divided five or more times. This is to reduce errors in the quantification of labelled cells with low CFSE fluorescence,
as is the case for five or more divisions, 
as described in Refs.~\cite{de2006estimating,ganusov}. Finally, the initial number of cells, $C_0$, is considered  a parameter to estimate in the model, since the actual number of transferred cells which made it to the lymph nodes or spleen is not measured.

We estimate model parameters with the ABC-SMC algorithm~\cite{toni}. Thus, the posterior distribution of the parameters is obtained by $T$ sequential applications of the ABC algorithm, where the posterior obtained in each iteration is used as prior for the next iteration. This algorithm requires the definition of prior distributions for the first iteration, a distance function,
a tolerance threshold for each iteration, and a perturbation kernel~\cite{toni}. 
We assume all parameters are initially distributed according to a uniform
prior distribution,
 as described in Table~\ref{priors}. When a prior distribution spans several orders of magnitude, the uniform distribution is taken over the exponent to efficiently explore the parameter space. Given the data point $x_D^g(t)$ which denotes the experimentally observed number of cells in generation $g$ at time $t$, for $g\in\{0,1,2,3,4,5+\}$, and the corresponding model prediction, $x^{g}_M(t)=M^g(t)$ for a particular choice of parameters ${\boldsymbol \theta}=(C_0,N_0,N,\lambda_0,\lambda,\alpha)$, the distance function is defined as \begin{equation}
    \label{distance}
        d(\text{model,data} \ | \ {\boldsymbol \theta}) = \sqrt{\sum_{g=0}^{G} \sum_{t \in \mathscr{T}} \left( \frac{x^{g}_M(t) - x_D^g(t)}{\sigma_D^g(t)} \right)^2},
\end{equation}
where $\mathscr{T}$ is the set of time points and depends on the clonotype of interest, $\sigma_D^g(t)$ represents the standard deviation of the experimental data point at time $t$ and generation $g$, and $G$ is the  merged generation class $G=5+$. In practice, we define the first tolerance threshold $\varepsilon_1$ in the ABC-SMC algorithm as the median value of the distances obtained from $10^4$ preliminary realisations, with the parameters sampled from the prior distributions in Table~\ref{priors}. The subsequent tolerance thresholds, $\varepsilon_j$, $j=2,\ldots,T$ can be then defined as the median of the distance values obtained from the previous iterations of the algorithm. Finally, we use a uniform perturbation kernel to perturb the parameters during the different iterations~\cite{toni}, and implement the algorithm for $T= 16$ in the case of the multi-stage model and $T=7$ for the single-stage one.

\begin{table}[h!]
    \centering
    \begin{tabular}{|c| l| l|}
    \hline
    Model parameters & Description & Prior distribution \\
    \hline
    $C_0$ & Initial number of cells & $C_0=10^x$, $x\sim U(4,6)$ \\
    $N_0$, $N$ & Number of stages & $U_{\text{discrete}}(1,50)$ \\
    $\lambda_0$, $\lambda$ & Birth rate & $\lambda_0=10^y$, $\lambda=10^z$, $y,z\sim U(-3,1)$  \\
    $\alpha$ & Death rate slope & $\alpha=10^w$, $w\sim U(-5,-1)$\\
    \hline
    \end{tabular}
\caption{Prior distributions for the model parameters.
Units for  $\lambda_0$, $\lambda$ and $\alpha$ are inverse hours ($h^{-1}$).}
\label{priors}
\end{table}

\begin{figure}[htp!]
    \centering
    \includegraphics[width=\textwidth]{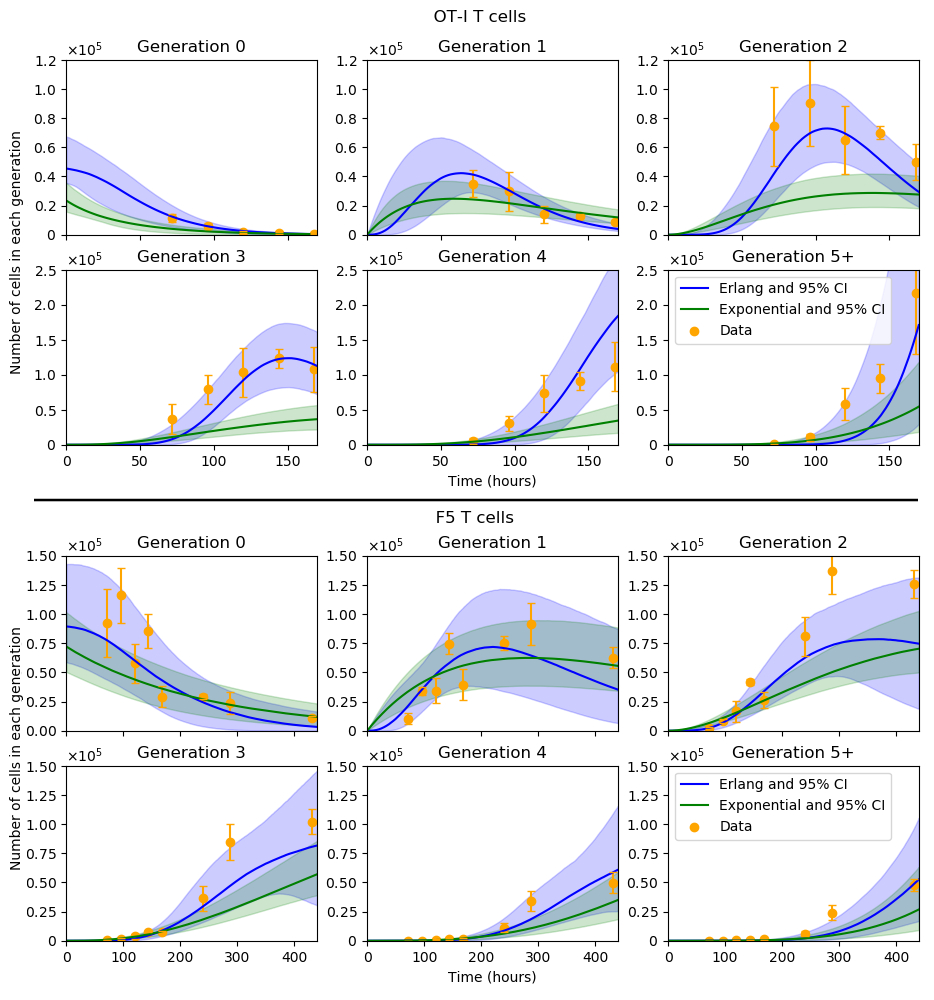}
    \caption{Exponential (solid green line) and multi-stage 
    (solid blue line)
    model predictions compared to the data sets (orange dots) for OT-I (upper panel) and F5 (lower panel) T~cells.
     Orange bars on the data points represent the standard deviation from the different  experimental replicates.
    The expected number of cells in each generation is plotted as a function of time. These predictions represent the median value of $10^4$ simulations with the accepted parameter values from the posterior distributions. Shaded areas represent 95$\%$ confidence intervals.}
    \label{pred_murine}
\end{figure}

The predictions obtained for each model, and for each clonotype (OT-I or
F5), are shown in Figure~\ref{pred_murine}. To obtain these predictions, we run the model with the parameters being sampled from the estimated posterior distributions and compute the median of all the simulations, which corresponds to the solid blue (multi-stage model) and green (exponential model) lines in Figure~\ref{pred_murine}. The bands surrounding the median predictions represent the 95$\%$ confidence intervals. The data points are plotted together with the standard deviation from the multiple experimental replicates. As shown in Figure~\ref{pred_murine}, the calibrated multi-stage model successfully captures the dynamics of the proliferating T~lymphocyte populations (OT-I and F5), whereas the single-stage model significantly underestimates the expected number of cells beyond generation $1$, particularly in the case of OT-I T~cells. Overall, the multi-stage model is able to explain the data from the OT-I transfer
experiment better, since this data set is less noisy than that of F5 T~cells. 

\begin{figure}[htp!]
    \centering
    \includegraphics[width=\textwidth]{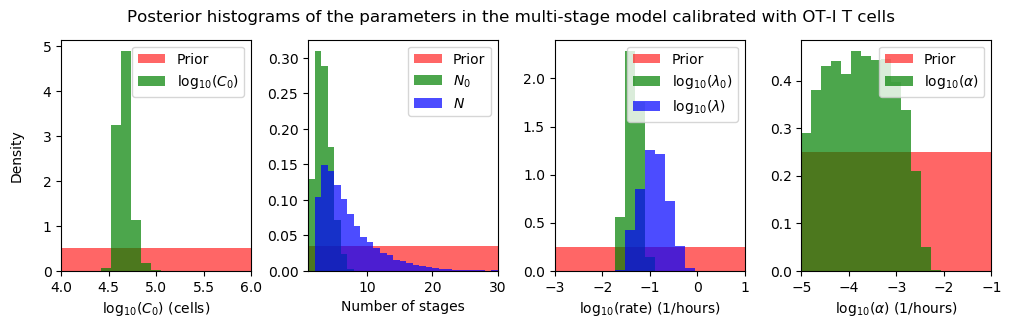}
    \includegraphics[scale=0.62]{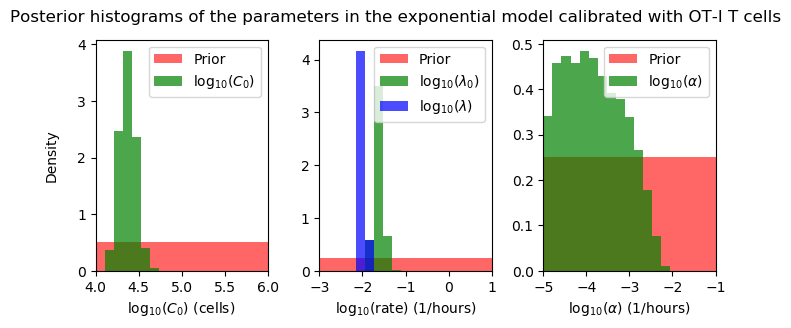}
    \caption{Posterior distributions (green and blue) for the parameters in the multi-stage ({\bf top}) and exponential ({\bf bottom}) model for OT-I T~cells. In the exponential model, the number of stages for all  generations is equal to 1, {\em i.e.,} $N_0 = N = 1$. Prior distributions are shown in red.}
    \label{post_OTI}
\end{figure}

\begin{figure}[htp!]
    \centering
    \includegraphics[width=\textwidth]{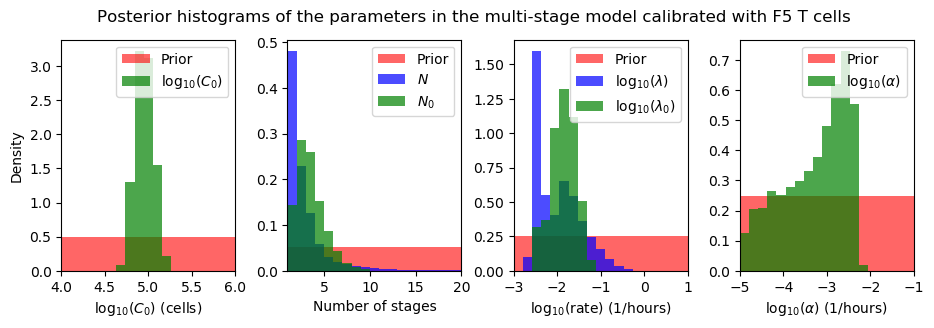}
    \includegraphics[scale=0.62]{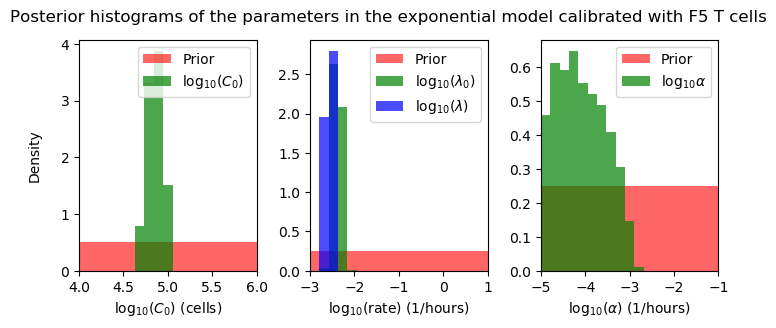}
    \caption{Posterior distributions (green and blue) for the parameters in the multi-stage ({\bf top}) and exponential ({\bf bottom}) model for F5 T~cells. In the exponential model, the number of stages for all  generations is equal to 1, {\em i.e.,} $N_0 = N = 1$. Prior distributions are shown in red.}
    \label{post_F5}
\end{figure}

\begin{figure}[htp!]
    \centering
    \includegraphics[width=0.85\textwidth]{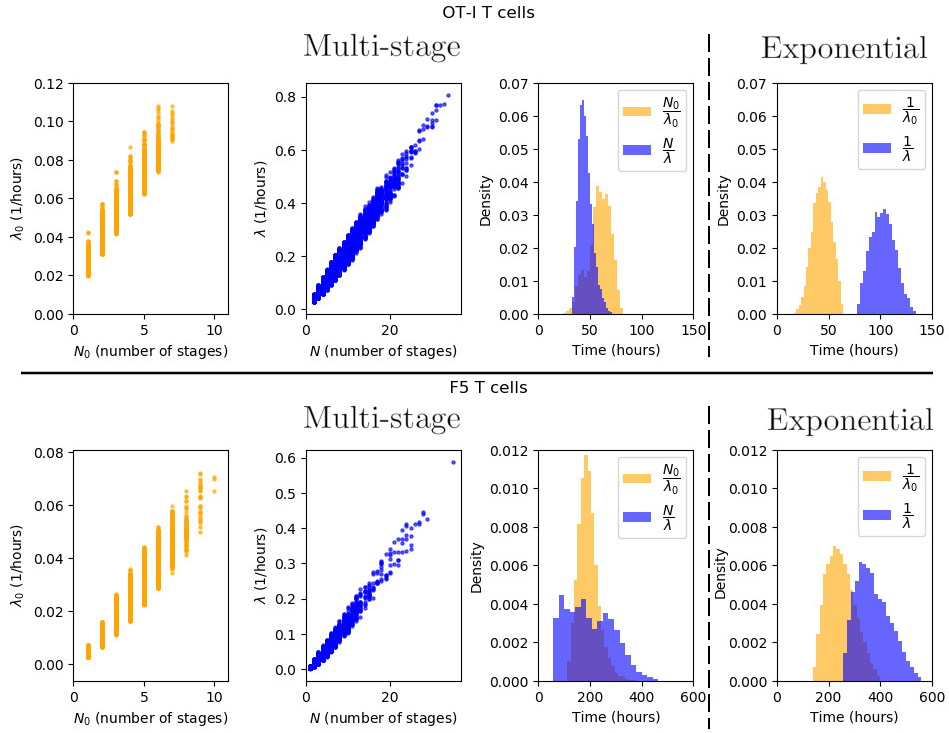}
    \caption{Joint posterior distributions ({\bf left}) of the number of stages $N_0$, $N$ and the birth rates $\lambda_0$, $\lambda$. Marginal posterior distributions ({\bf right}) for the mean time to first and subsequent divisions estimated from the multi-stage model (third column) and the exponential model (fourth column).}
    \label{div_time_mice}
\end{figure}

The marginal posterior distributions for each parameter are shown in green and blue in Figures~\ref{post_OTI} and \ref{post_F5}, for the multi-stage and exponential models, and the (uniform) prior distributions are plotted in red. Summary statistics for these posterior distributions are reported in Tables~\ref{summ_stat_erlang_OTI} and \ref{summ_stat_erlang_F5}. Cell death is governed by the parameter $\alpha$, and is estimated to be low for both models and clonotypes, suggesting that cell death does not have a significant impact on the dynamics during lymphopenia, which is in fact dominated by cell division. This result is in agreement with Hogan {\em et al.}~\cite{hogan}, where the death rate is assumed to be zero. The initial number of cells can be estimated with relative success, and does not seem to depend heavily on the model considered. On the other hand, cell division is governed by parameters $(N_0,\lambda_0,N,\lambda)$, with $N_0=N=1$ in the exponential model. We note that in both models, $N_0/\lambda_0$ and $N/\lambda$ represent the mean time until the first and subsequent divisions, respectively. Although all division-related parameters can be estimated from the data, for both models and clonotypes, a correlation between the division rate and the number of stages is seen in the scatter plots of Figure~\ref{div_time_mice}. Instead of plotting the marginal posterior distributions for these parameters, one can consider the posterior distribution for the mean times $N_0/\lambda_0$ and $N/\lambda$ (see Figure~\ref{div_time_mice}). The fact that $N=1$ is never chosen as an accepted parameter value in the posterior distribution for the multi-stage model and the OT-I clonotype already suggests that a multi-stage representation of cell division is preferred for this clonotype. On the other hand  for the F5 clonotype the marginal distribution for $N$ shows a 
non-zero frequency for the value $1$, but larger values of $N$ are also represented in its posterior distribution. The mean time to both first and subsequent divisions, $N_0/\lambda_0$ and $N / \lambda$, are significantly longer for the F5 clonotype than the OT-I clonotype. In fact, our results estimate that F5 T~cells divide slowly compared to OT-I cells, requiring on average $192$ hours to carry out a first division  ($59$ hours taken by OT-I T~cells), as depicted in Figure~\ref{div_time_mice} for the multi-stage model.
The time to subsequent divisions is represented by the blue histograms. Interestingly, our estimation of the mean time to first division of OT-I cells, on average 59 hours, is close to the value obtained in Ref.~\cite{hogan}  (52 hours when considering the best fit parameter estimates). 
In the case of F5 cells, we predict an average of 192 hours
 to undergo their first division, whereas
 Hogan {\em et al.} obtained a value of 137 hours.
We  note that the value $137$ hours is within the range covered by our predicted posterior distribution.

\begin{table}[htp!]
    \centering
    \begin{tabular}{|c|c|c|c|c|c|}
    \hline
    Parameter & Minimum & Maximum & Mean & Median & Standard deviation \\
    \hline
    $C_0$ & $3.00 \cdot 10^4$ & $1.03 \cdot 10^5$ & $4.67 \cdot 10^4$ & $4.54 \cdot 10^4$ & $8.03 \cdot 10^3$ \\
    \hline
    $N_0$ & 1 & 7 & 2.83 & 3 & 1.23 \\
    \hline
    $N$ & 2 & 34 & 6.59 & 5 & 4.30 \\
    \hline
    $\lambda_0$ & $1.98 \cdot 10^{-2}$ & $1.08 \cdot 10^{-1}$ & $4.64 \cdot 10^{-2}$ & $4.56 \cdot 10^{-2}$ & $1.45 \cdot 10^{-2}$ \\
    \hline
    $\lambda$ & $2.80 \cdot 10^{-2}$ & $8.08 \cdot 10^{-1}$ & $1.48 \cdot 10^{-1}$ & $1.20 \cdot 10^{-1}$ & $1.01 \cdot 10^{-1}$ \\
    \hline
    $\alpha$ & $1.00 \cdot 10^{-5}$ & $5.97 \cdot 10^{-3}$ & $5.06 \cdot 10^{-4}$ & $1.76 \cdot 10^{-4}$ & $7.47 \cdot 10^{-4}$ \\
    \hline
    \end{tabular}
\caption{Summary statistics for the posterior distributions of the multi-stage model for the OT-I clonotype.}
\label{summ_stat_erlang_OTI}
\end{table}

\begin{table}[htp!]
    \centering
    \begin{tabular}{|c|c|c|c|c|c|}
    \hline
    Parameter & Minimum & Maximum & Mean & Median & Standard deviation \\
    \hline
    $C_0$ & $4.74 \cdot 10^4$ & $1.85 \cdot 10^5$ & $9.26 \cdot 10^4$ & $8.94 \cdot 10^4$ & $2.22 \cdot 10^4$ \\
    \hline
    $N_0$ & 1 & 10 & 3.01 & 3 & 1.53 \\
    \hline
    $N$ & 1 & 35 & 2.42 & 2 & 2.57 \\
    \hline
    $\lambda_0$ & $2.68 \cdot 10^{-3}$ & $7.20 \cdot 10^{-2}$ & $1.70 \cdot 10^{-2}$ & $1.47 \cdot 10^{-2}$ & $1.07 \cdot 10^{-2}$ \\
    \hline
    $\lambda$ & $2.06 \cdot 10^{-3}$ & $5.88 \cdot 10^{-1}$ & $2.20 \cdot 10^{-2}$ & $9.54 \cdot 10^{-3}$ & $3.90 \cdot 10^{-2}$ \\
    \hline
    $\alpha$ & $1.00 \cdot 10^{-5}$ & $6.21 \cdot 10^{-3}$ & $1.35 \cdot 10^{-3}$ & $8.19 \cdot 10^{-4}$ & $1.40 \cdot 10^{-3}$ \\
    \hline
    \end{tabular}
\caption{Summary statistics for the posterior distributions of the multi-stage model for the F5 clonotype.}
\label{summ_stat_erlang_F5}
\end{table}

Together, our multi-stage model results indicate that 
OT-I T~lymphocytes
require on average 59 hours for their first division, and a bit less
time, 46 hours, for sub
 subsequent divisions (see upper left plot of Figure~\ref{div_time_mice}). 
 Our results allow us to conclude that a multi-stage model, 
 with a constant division rate after the first division event, is
 a suitable description of  the dynamics
 of recovery from lymphopenia~\cite{gett2000cellular}. On the other hand, the multi-stage model estimates that F5 cells take on average slightly less than 200 hours to divide, both for the first or subsequent division rounds, as shown in the lower left plot of Figure~\ref{div_time_mice}. This difference might be related to the different response of OT-I and F5 T~cells to lymphopenia~\cite{hogan}. 
 Finally, it is evident that for both clonotypes the exponential model 
 (see Figure~\ref{div_time_mice}) found a shorter time to first division than to subsequent ones, contradicting previous findings~\cite{hawkins2007model,kinjyo2015real,markham2010minimum}. This seems directly related to the fact that, overall, the exponential model is not able to capture the observed cell dynamics for neither of the clonotypes, as can be observed in Figure~\ref{pred_murine}.

\section{Conclusion}  
\label{discussion}

The multi-stage model implemented here takes cell death into account
while retaining practical advantages: we are able to find closed
expressions for the mean number of cells in each generation, and
generate numerical realisations using the Gillespie algorithm \cite{gillespie1,gillespie2}. A
longer mean time to first division, $N_0/\lambda_0$, than mean
time to subsequent divisions, $N/\lambda$, is a natural part of the
framework without the need to introduce extra parameters. On the other
hand, our calculations rely on the assumption that cells are
independent of each other. Further work is needed to model the late
stages of lymphopenia-induced proliferation, or possible
cell fate correlations within family trees. \\

{\bf Acknowledgements}
This work has been supported by the European Commission through the Marie Skłodowska-Curie Action (H2020-MSCA-ITN-2017) -Innovative Training Network- Quantitative T-cell Immunology and Immunotherapy (QuanTII), project number 764698. We also acknowledge the University of Leeds for the permission to use the High Performance Computing facilities ARC1 and ARC2.


\end{document}